\documentclass[useAMS,usenatbib,psfig,letter]{mn2e}
\usepackage{graphicx}
\begin{document}
\title[Circumstellar material around Nova Oph 2015]{Flash-ionization of
pre-existing circumstellar material around Nova Oph 2015}
\author[Munari and Walter]{U. Munari$^{1}$ and F. M. Walter$^{2}$\\
$^{1}$INAF Astronomical Observatory of Padova, 36012 Asiago (VI), Italy\\
$^{2}$Dept. of Physics and Astronomy, Stony Brook University, Stony Brook, NY 11794-3800, USA}

\date{Accepted .... Received ....; in original form ....}

\pagerange{\pageref{firstpage}--\pageref{lastpage}} \pubyear{2010}

\maketitle

\label{firstpage}

\begin{abstract}
We have obtained daily high resolution Echelle spectroscopy of Nova Oph 2015
during its initial evolution.  It reveals the presence of pre-existing
circumstellar material around the nova, which is best interpreted as the
wind of an evolved companion.  On earliest observations, the emission line
profiles of Nova Oph 2015 displayed a very narrow emission component
(FWHM$\sim$60 km sec$^{-1}$), recombining over a time scale of 5 days and
showing constant low velocity (RV$_\odot$=$-$4.5 km sec$^{-1}$), that we
interpret as coming from the wind of the secondary recombining after the
ionization from the initial UV-flash of the nova.  The underlying broad
component underwent a marked reduction in both FWHM and width at zero
intensity (the latter declining from 4000 to 2000 km sec$^{-1}$ in ten days)
while increasing by 6$\times$ in flux, that we believe is the result of the
high velocity ejecta of the nova being slowed down while trying to expand
within the surrounding wind of the companion.  Novae with evolved
secondaries are very rare in the Galaxy, amounting to $\sim$3\% of the total
according to recent estimates.  Among them Nova Oph 2015 is perhaps unique
in having displayed a long rise to maximum brightness and a slow decline from
it, a FeII-type classification (contrary to prevailing He/N) and a probable
sub-giant luminosity class for the secondary (instead of the giant (e.g.  RS
Oph) or supergiant (e.g.  V407 Cyg) class for the others).
\end{abstract}
\begin{keywords}
Stars: novae
\end{keywords}

\section{Introduction}

Nova Oph 2015 was discovered by Y. Sakurai (Japan) on Mar 29.766 UT at
unfiltered $\sim$12.2 magnitude (cf CBET 4086) and received the designation
PNV J17291350-1846120 when posted at the Central Bureau for Astronomical
Telegrams's TOCP webpage.  Spectroscopic observations on Mar 30 by Fujii
(2015) and Ayani (2015), and on Apr 2 by Danilet et al.  (2015) showed
the new transient to be a nova.  Their classification was ``He/N" type. 
However, at the time of these early spectroscopic observations the nova was
still $\sim$3 mag below maximum brightness and spectroscopic observations
obtained around maximum on Apr 11 by Munari et al.  (2015) revealed instead
a textbook example of a ``FeII" type nova dominated by a reddened and strong
continuum, with intense emission lines from Balmer, FeII, CaII and OI, all
showing deep P-Cyg absorptions.  None of the HeI, NII and NIII emission
lines distinctive of an `He/N" type was present.  

During their initial rise in brightness, the expanding photospheres of FeII
novae cool from the initial extremely high temperatures (at the time when
the electron degeneracy is lifted) to the $\sim$8000 K characteristic of
maximum optical brightness, and in doing this they have to pass through the
hotter temperatures characteristic of He/N-type spectra (Seitter 1990),
which seems a plausible explanation for the apparent conflict in the
spectral classifications of Nova Oph 2015.  This should not be confused with
the rare and still unclear phenomenon of {\it hybrid}-novae (Williams 1992),
where features typical of {\it both} He/N and FeII types are simultaneously
present and independently evolve during the {\it post-maximum} decline.  A
FeII classification also fitted the IR observations of Nova Oph 2015 by
Banerjee at al.  (2015), although they noted that for a brief period HeI
lines were stronger than usual for this type of nova.  Here it is worth
reminding that the FeII and He/N classification scheme introduced by
Williams (1992) is applicable to novae only around maximum brightness and
early decline from it.

Classical novae are compact binaries, typically of a few hours orbital
period, wherein a low-mass unevolved star that fills its Roche lobe
transfers material to a more massive white dwarf companion.  The transfer
through L1 occurs at a low rate and little matter is dispersed into the
circumstellar environment (Warner 1995, Hellier 2001).  The fast evolving
thermonuclear runaway that initiates the nova outburst increases (in a
matter of minutes) the effective temperature of the white dwarf envelope to
several 10$^5$ K (Starrfield et al.  2008), before the violent expansion
following the lifting of the electron degeneracy drives it through a rapid
cooling.  The short-lived surge in effective temperature produces a powerful
UV flash that disperses into the empty circumstellar environment emptiness
and goes undetected; the nova will be discovered only a few days later when
the optically thick ejecta will have expanded enough to become a bright
optical source.

There is however a very rare type of nova ($\sim$3\% of the total, Williams
et al.  2014) in which the initial UV flash interacts with dense
pre-existing circumstellar material, the prototypes being the recurrent
novae RS Oph and V407 Cyg (Evans et al.  2008; Munari et al.  2011).  In
such systems, the WD orbits - with periods of a few years - within the
extended wind of a late type giant companion.  Such a wind is able to
completely absorb the initial UV flash, become largely ionized, and shine
for several days during the following recombination.  In this letter we
present evidence that Nova Oph 2015 is a new member of this highly exclusive
club of novae with evolved secondaries.  In some equally rare novae, the
initial UV flash manifested itself at much later times.  In Nova Vul 2007,
about one year past eruption, the UV flash reached and ionized gas blown off
by the WD progenitor about 14\,000 yrs earlier (Wesson et al.  2008). 
Several months after the 2011 outburst of the recurrent Nova T Pyx, the UV
flash reached and ionized blobs of material ejected during the previous
outburst (Shara et al.  2015).

  \begin{figure}
     \centering
     \includegraphics[width=8.5cm]{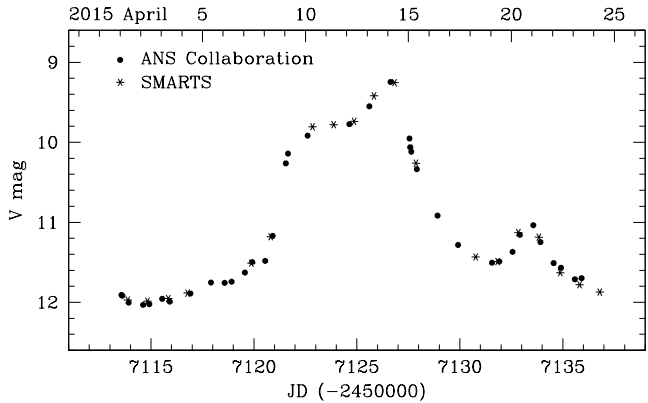}
     \caption{$V$-band photometric evolution of Nova Oph 2015, covering
     the rise to maximum and initial decline.} 
     \label{fig1}
  \end{figure}

  \begin{figure}
     \centering
     \includegraphics[width=8.5cm]{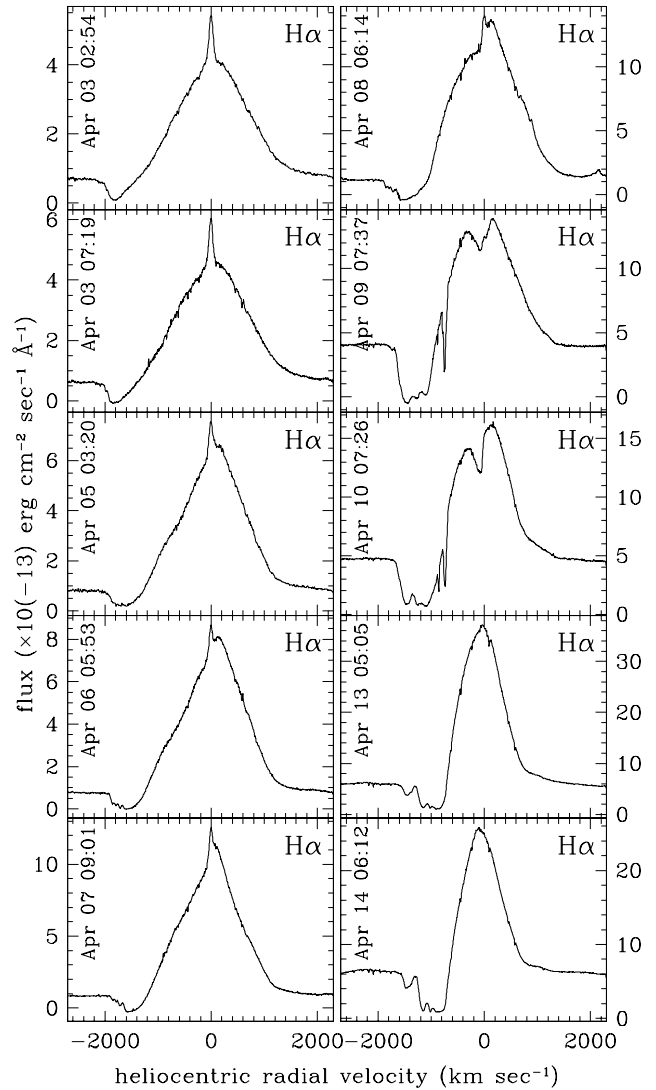}
     \caption{H$\alpha$ profile evolution of Nova Oph 2015, covering
     the rise to maximum (occurring on Apr 14.2 UT). The UT date of
     each spectrum is given.}
     \label{fig2}
  \end{figure}

  \begin{figure}
     \centering
     \includegraphics[width=8.5cm]{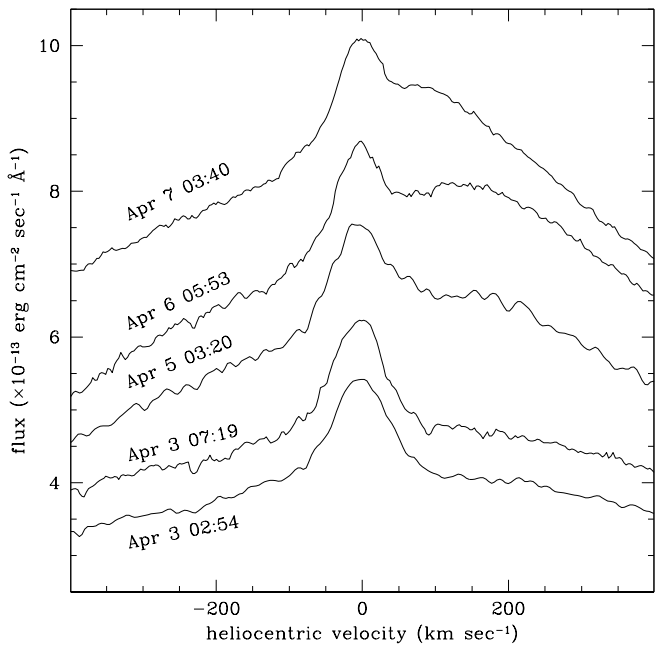}
     \caption{Expanded view of some of the H$\alpha$ profiles from
     Figure~2 to highlight the narrow component superimposed to the
     broader one. An offset in flux (-2.7 units) has been applied to 
     Apr 7 03:40 spectrum to better fit the figure.}
     \label{fig3}
  \end{figure}

\section{Observations}

High-resolution spectroscopic observations of Nova Oph 2015 were obtained
with telescopes SMARTS 1.5m + CHIRON from CTIO (Chile), with 1.82m + REOSC
Echelle from Asiago and with 0.61m + Astrolight Instruments mk.III
Multi-Mode Spectrograph from Varese.  Given the low target altitude above
the horizon for observation from Italy, the spectra from Asiago and Varese
were obtained with a 2-arcsec wide and 20-arcsec long slit aligned along
the parallactic angle, while SMARTS 1.5m telescope uses a 1.5-arcsec
diameter fiber to feed light to CHIRON spectrograph (Tovokin et al. 2013).

Optical photometry of Nova Oph 2015 was obtained with ($i$) ANS
Collaboration telescopes 30 (Cembra, Italy) and 210 (Atacama, Chile) and
reduced against a local photometric sequence extracted from all-sky APASS
survey (Henden et al.  2012, Munari et al.  2014), and ($ii$) with SMARTS
1.3-m + ANDICAM from CTIO (Chile), and reduced against nightly observations
of all-sky standard stars (Walter et al. 2012).

\section{Emission line profiles}

The photometric evolution of Nova Oph 2015 during it rise toward maximum
brightness is shown in Figure~1.  The rise was slow, with a pre-maximum
halt lasting a week while the nova was 2.5/3.0 mag below maximum.  The
maximum was reached at $V$$\sim$9.25 mag around Apr 14.2 UT (2457126.7), two
weeks past initial discovery.

The evolution of the H$\alpha$ line profile during the rise toward maximum
is illustrated in Figure~2 (H$\beta$ and H$\gamma$ evolved similarly). 
Apart from the complex system of P-Cyg absorptions typical of FeII novae,
the profile evolution is characterized by ($i$) the initial presence of a
very narrow component, and ($ii$) the monotonic reduction in width of the
underlying broad component, in particoular the gradual suppression of the
high velocity wings.

\subsection{The narrow component}

A narrow component has been seen during the earliest phases of a nova
outburst only for those erupted in symbiotic binary system, like RS Oph and
V407 Cyg, where the WD is engulfed by the wind of the late type giant
companion.  This narrow component should not be confused with a similar
feature observed in a small group of novae (KT Eri, YY Dor, LMC 1990b, LMC
2009a, V394 Cra, U Sco and DE Cir) much later in the evolution and that is
belived to originate either from projection effect of bipolar ejecta or from
the unveiling of the central binary that has resumed accretion (Walter et
al.  2012, Mason \& Walter 2013, Shore at al.  2013, Mason \& Munari 2014,
Munari, Mason, \& Valisa 2014).

The narrow component is visible in Nova Oph 2015 for the first six days of
our monitoring (cf Figure~2), during which its heliocentric radial velocity
remains constant within 2 km sec$^{-1}$ (essentially the measurement error)
of the average value $-$4.5 km sec$^{-1}$.  The narrow component has a
Gaussian profile (cf Figure~3) with a FWHM $\sim$60 km sec$^{-1}$.  The
evolution with time of the integrated flux of the narrow component is
plotted Figure~4 (top panel).  It shows a steady decline, with a 1/$e$
recombination time scale of $t_{\rm rec}$=5 days.  In analogy with RS Oph
and V407 Cyg, we interpret the narrow component as originating from the wind
of the secondary recombining after the sudden ionization caused by the initial
UV-flash.  In this context, the FWHM of 60 kms sec$^{-1}$ nicely fits the
expectation from the slow wind of an evolved star and the $-$4.5 km
sec$^{-1}$ heliocentric radial velocity corresponds to that of the cool
giant (combining galactic and orbital motions).  The recombination time
scale (in hours) is related to electronic density ($n_{\rm e}$) and
temperature ($T_{\rm e}$) by
\begin{equation}
t_{\rm rec} = 1.15 \left(\frac{T_{\rm e}}{10^4 {\rm ~K}}\right)^{0.8}
\left(\frac{n_{\rm e}}{10^9 {\rm ~cm}^{-3}}\right)^{-1} 
\end{equation}
(Ferland 1997), to which correspond a density of $n_{\rm e}$=1$\times$10$^7$
cm$^{-3}$ for a typical $T_{\rm e}$=1$\times$10$^4$~K, a value of $n_{\rm
e}$ similar to that derived for the winds in RS Oph and V407 Cyg. 

A major difference with RS Oph (Skopal et al.  2008) and V407 Cyg (Munari et
al.  2011), is the lack of an even sharper absorption superimposed to the
narrow component.  This very sharp absorption (FWHM$\sim$15 km sec$^{-1}$)
originates is the outer neutral portion of the wind not reached by the
initial UV-flash, which is completely absorbed by the gas inner to it.  The
lack of an external neutral zone suggests that the wind of the companion in
Nova Oph 2015 extends much less than in RS Oph or V407 Cyg, i.e.  the
evolved star in Nova Oph 2015 is of lower luminosity and/or earlier spectral
type.  This argument is reinforced by noting that the eruption of Nova Oph
2015, of FeII type, was probably much less energetic that those of RS Oph 
and V407 Cyg, both of the He/N type.

\subsection{The broad component}

The evolution of the width and the integrated flux of the broad component of
H$\alpha$ is shown in Figure~4.  It is characterized by a continuous
sharpening of the profile and a parallel continuous increase of its
integrated flux.  In principle, the sharpening could be the result of the
ejecta's pseudo-photosphere cooling because of the expansion: the decreasing
number of emitted ionizing photons is unable to reach and ionize more
distant - and therefore faster moving - ejecta.  This scenario requires
however a parallel {\it reduction} in the integrated flux because of the
decreasing amount of ionized gas.  The very fact that the broad component of
H$\alpha$ instead {\it increased} its flux by 6$\times$ during the
sharpening period (cf Figure~4), argues for the latter being caused by a
deceleration of the expanding ejecta.  This would be caused by the ejecta's
expansion within the wind of the evolved companion, with the resulting shock
sustaining the ionization of the swept up material.  The measurement of the
FWHM of the broad component is grossly perturbed by the P-Cyg absorption
components on the blue side, so we have also measured the velocity at zero
intensity of the red wing of H$\alpha$ (unperturbed by absorptions), and
plotted it as {\it ZI-red} on the bottom panel of Figure~4.

The ZI-red velocity halves, from $\sim$2000 to $\sim$1000 km~sec$^{-1}$,
during the 11 days of pre-maximum phase.  The ejecta first slamming onto the
slow-moving pre-existing wind are those moving at higher velocities and
responsible for the emission in the wings of H$\alpha$ profile, which are
rapidly suppressed, as clearly illustrated in Figure~2.  A closely similar
profile-sharpening and wing-suppression was observed in V407 Cyg, where
H$\alpha$ took 5 days to half its FWHM (Munari et al.  2011).  After maximum
brightness, the ZI-red velocity of H$\alpha$ did not change significantly,
remaining close to $\sim$1100 km~sec$^{-1}$ suggesting that no further
deceleration took place because the ejecta broke free of the wind and
continued their expansion in the surrounding space.  The fact that the
sharpening of H$\alpha$ continued for months in V407 Cyg, supports the
notion that the wind of its secondary extended over larger distances and
carried a larger mass than for Nova Oph 2015.

  \begin{figure}
     \centering
     \includegraphics[width=8.5cm]{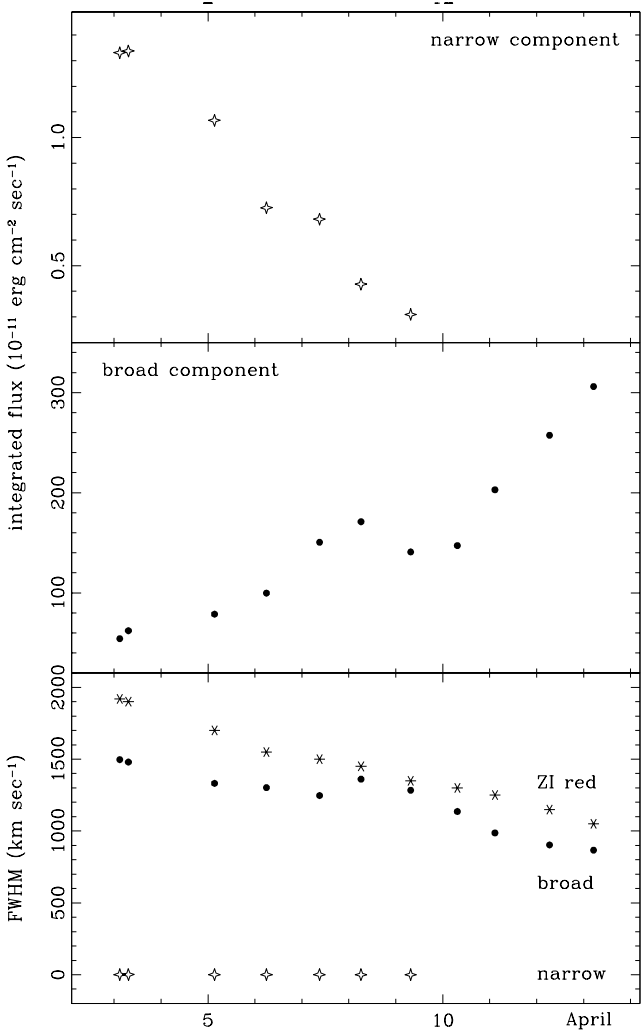}
     \caption{Evolution in integrated flux and FWHM for the H$\alpha$ narrow 
     and broad components. `ZI red' is the velocity at zero intensity of the
     red wing of H$\alpha$, unperturbed by the P-Cyg absorptions that plague
     the blue one.}
     \label{fig4}
  \end{figure}

\section{The nature of the cool companion}

At the astrometric position of the nova (A.  Henden, private communication)
$\alpha$=17:29:13.416 $\delta$=$-$18:46:13.80 ($\pm$0.030 arcsec on both
axes), corresponding to galactic coordinates $l$=6$^\circ$.6426,
$b$=8$^\circ$.5757, there is no entry in neither the 2MASS or AllWISE
infrared catalogs, nor in GSC, PPMX or Nomad optical ones.  They all list a
few sources within some arcsec from the nova, but they are all distinct from
the nova as proved by direct imaging.  

The total Galactic extinction in the direction of the nova is $A_J$=0.51,
$A_H$=0.33, and $A_K$=0.21, averaging the values from Schlegel et al. 
(1998) and Schlafly \& Finkbeiner (2011) maps.  This agrees with the
reddening derived from the diffuse interstellar band at 6614 \AA, for which
we have measured an equivalent width of 0.118 \AA\ on our high resolution
spectra.  This corresponds to a reddening of $E_{B-V}$=0.52 following the
calibration by Munari (2014) or $E_{B-V}$=0.56 from Kos and Zwitter (2013). 
For the standard $R_V$=3.1 extinction law, the averaged $E_{B-V}$=0.54
transforms into $A_K$=0.21, the same value as above.  At the high Galactic
latitude of the nova, the line of sight exits the Galactic Thin Disk $\sim$1
kpc from the Sun (having accumulated up to that point $E_{B-V}$$\sim$0.25
for standard galactic extinction models), travels through some relatively
empty space and then reaches the Bulge.  A partnership with the Bulge is far
more probable for two basic reasons: ($i$) the nova seems too faint (the
extinction corrected value at maximum is $V^{\rm max}_{\circ}$$\sim$7.6) for
a distance within the Galactic disk, and ($ii$) the extinction measured for
the nova is too large to be generated within the Galactic disk alone and
instead perfectly matches the total Galactic extinction along the line of
sight.  A location within the Galactic Bulge would result in an absolute
$M_V$=$-$7.1 mag, a reasonable value for the moderately slow decline-rate
displayed by this nova (Downes \& Duerbeck 2000).  It is worth noticing that
the statistics of the nova population in the M31 Andromeda galaxy place the
large majority of them within its bulge (Shafter \& Irby 2001, Williams et
al.  2014).

The limits for completeness of 2MASS detections around the nova are
$J$$\sim$16.6, $H$$\sim$15.6 and $K$$\sim$15.2, as derived by inspection of
the histograms of stellar counts vs magnitude for sources within 5 arcmin of
the nova.  Under an extinction of $A_K$=0.21, an early-K giant located in
the Galactic Bulge would shine at $K$$\sim$13, an early M giant at
$K$$\sim$10.5, and a late M giant at $K$$\sim$7.5 (Koornneef 1983, Frogel \&
Whitford 1987, Sowell et al.  2007), amply within the completeness limit for
2MASS.  This excludes a giant as the donor star in Nova Oph 2015, while
leaves open the possibility for a sub-giant.  The wind of a sub-giant should
extend much less than for a giant, which agrees well with our scenario that
the circumstellar medium around Nova Oph 2015 was overrun by the nova ejecta
much earlier than for RS Oph and V407 Cyg.  

\section{Discussion}

Nova Oph 2015 is probably the first time that an FeII-type eruption is
seen in a system containing an evolved secondary. In the other cases, like RS Oph
and V407 Cyg, the nova eruptions have been of the He/N-type. An FeII-type 
for Nova Oph 2015 nicely agrees with its partnership to the Galactic Bulge,
where FeII-types are almost exclusively observed while He/N-types are
generally confined to the Galactic Disk (Della Valle \& Livio 1998).

The He/N type novae with evolved companions are characterized by such a
rapid rise to maximum that it is easy to miss observationally.
Nova Oph 2015 has been instead discovered two weeks before optical maximum. 
Being discovered several days before reaching maximum is ordinary for FeII
novae, a period of time spent in completing the rise in brightness with
usually a brief pause termed ``pre-maximum halt" (McLaughlin 1960).  What is
not ordinary is the fact that observed portion of the pre-maximum halt in
Nova Oph 2015 lasted $\sim$8 days, longer than traditionally observed
(Payne-Gaposchkin 1964, Warner 1995, Hounsell et al.  2010).  Such a long
lasting pre-maximum plateau could be the result of an apparent balance
between the decline in brightness of the recombining wind (after the
ionization by the initial UV-flash) and the rise in brightness of the
expanding nova ejecta.

A distinguishing feature of Nova Oph 2015 is the luminosity classification
of its secondary star: not a giant or supergiant (no 2MASS detection), and
unlikely to be a main sequence (because of the presence of a thick and slow
wind), which leaves open only the possibility that it is a sub-giant.  A
star evolves much faster through the sub-giant period than during the
following giant phase and loses mass at a much lower rate, which would
explain the paucity of known WD + sub-giants among novae.  This poses the
question on how the mass is transferred from the sub-giant to the WD: by
capture from its wind or via Roche lobe overflow through L1 ?  The radius of
a star evolving through the sub-giant phase rapidly expands from that of a
main sequence to that of a giant.  This means that, for a given orbital
separation, it would match the requirement for Roche-lobe filling only
briefly.  Before that the system would have been probably dormant, a main
sequence star being unable to transfer mass via wind to the WD companion. 
After that, the expansion in radius of the secondary would bring the binary
system under common-envelope conditions.

Given its rarity, it will be important to assess the evolutionary status of
the secondary star of Nova Oph 2015.  It could be derived by post-outburst
IR observations (deeper than 2MASS), when the system will have returned to
quiescence conditions, with the results compared to theoretical isochrones. 
If protracted over time, these IR observations would allow to search for
ellipsoidal modulation of the lightcurve betraying a Roche-lobe filling
secondary, in which case the orbital period may be expected to be (much)
shorter than in RS Oph (456 days) and of the order of some weeks to a few
months.

\section{Acknowledgements}

We are grateful to J. Hambsch, A. Frigo, S. Dallaporta, P. Valisa, J. Kos,
A.  Siviero and M. Zerial that in various ways helped with some of the
observations here discussed.

\bsp

\label{lastpage}

\end{document}